\documentclass[fleqn,usenatbib,usedcolumn]{mnras}
\makeatletter\ifdefined\@xvipt\def\@versiontest{3.0}
  \ifx\@version\@versiontest
    \def\@version{3.0 {\tiny (the 2020 one!)}}
  \fi
\fi\makeatother
\usepackage[british]{babel}             
\usepackage{newtxtext}                  
\usepackage[slantedGreek]{newtxmath}    
\usepackage[T1]{fontenc}                
\usepackage{graphicx}                   
\hypersetup{pdfauthor={D. A. Green and N. Madhusudhan},
            pdftitle={Search for radio emission from the exoplanets
              Qatar-1b and WASP-80b near 150 MHz using the Giant
              Metrewave Radio Telescope},
            pdfkeywords={radio continuum: planetary systems,
              planets and satellites: individual: Qatar-1b,
              planets and satellites: individual: WASP-80b},
            bookmarksnumbered=true}     
\title[Search for radio emission from the exoplanets]{Search for radio
emission from the exoplanets Qatar-1b and WASP-80b near 150~MHz using
the Giant Metrewave Radio Telescope}

\author[Green \& Madhusudhan]{D.~A.~Green,$^1$\thanks{email:
        {\tt dag@mrao.cam.ac.uk}} N.~Madhusudhan$^2$\\
        $^1$Astrophysics Group, Cavendish Laboratory,
           19 J.~J.~Thomson Avenue, Cambridge CB3 0HE, United Kingdom\\
        $^2$Institute of Astronomy, Madingley Road,
           Cambridge CB3 0HA, United Kingdom}

\date{Accepted 2020 October 12. Received 2020 October 8; in original
form 2020 September 25}

\volume{{\rm in press}}

\pubyear{2020}

\begin{document}
\label{firstpage}

\pagerange{\pageref{firstpage}--\pageref{lastpage}}

\maketitle

\begin{abstract}
We present radio observations made towards the exoplanets Qatar-1b and
WASP-80b near 150~MHz with the Giant Meterwave Radio Telescope. These
targets are relatively nearby irradiated giant exoplanets, a hot Jupiter
and a hot Saturn, with sizes comparable to Jupiter but different masses
and lower densities. Both the targets are expected to host extended H/He
envelopes like Jupiter, with comparable or larger magnetic moments. No
radio emission was detected from these exoplanets, with $3\sigma$ limits
of 5.9 and 5.2~mJy for Qatar-1b and WASP-80b, respectively, from these
targeted observations. These are considerably deeper limits than those
available for exoplanets from wide field surveys at similar frequencies.
We also present archival VLA observations of a previously
reported radio source close to 61 Vir (which has three exoplanets).
The VLA observations resolve the source, which we identify as an
extragalactic radio source, i.e.\ a chance association with 61 Vir.
Additionally, we cross-match a recent exoplanet catalogue with the
TIFR GMRT Sky Survey ADR1 radio catalogue, but do not find any convincing
associations.
\end{abstract}


\section{Introduction}\label{s:introduction}

Thousands of exoplanets are known today and transit surveys are expected to
detect thousands more in the near future. The exoplanets detected to date span
a diverse range of orbital and bulk properties and host stars. We are now
entering a new era where detailed observations are characterising their
atmospheric processes and chemical compositions (e.g.\
\citealt{2017JGRE..122...53D, 2018haex.bookE.100K, 2019ARA&A..57..617M}), from
observations made at ultraviolet to infrared wavelengths. Any detection of
radio emission from exoplanets would present a unique opportunity to
characterise their magnetic processes and internal structures which are
inaccessible from other observations. Solar system planets span diverse
intrinsic magnetic field strengths, from 0.2~G in the ice giants,
through 0.5~G in the Earth, to 4.2~G in Jupiter, suggesting the presence
of dynamos in their convective interiors \citep{2003E&PSL.208....1S}.
Theoretical studies have predicted that cyclotron radio emission from
giant exoplanets with magnetic field strengths comparable to Jupiter's
are potentially observable with existing and upcoming radio facilities
(e.g.\ \citealt{2004ApJ...612..511L, 2007A&A...475..359G,
2001Ap&SS.277..293Z, 2007P&SS...55..598Z}). For close-in giant
exoplanets, the dominant source of energetic electrons interacting with
the planetary magnetic field is generally the stellar wind (e.g.\
\citealt{2007P&SS...55..598Z}), but the effect of auroral processes and
potential exomoons have also been investigated (e.g.\
\citealt{2011MNRAS.414.2125N, 2012MNRAS.427L..75N,
2014ApJ...791...25N}).

The gyrofrequency ($f_{\rm cy}$) for the electron--cyclotron maser
radiation due to a planetary magnetic field ($B_{\rm p}$) is given by
$f_{\rm cy} = 2.8 (B_{\rm p} / 1~{\rm G})$~MHz, implying that emission
from exoplanets with $B_{\rm p} \sim 1{-}100$~G can be detected in the
frequency range of $2.8{-}280$~MHz. For Jupiter, the observed radio
frequency cut-off in its cyclotron emission leads to an estimated
maximum $B_{\rm p}$ of 14~G. While future facilities such as the Square
Kilometre Array (SKA) will be able to detect Jovian and sub-Jovian
magnetic fields, exoplanets with $B_{\rm p}$ of a few times the Jovian
value (e.g.\ $\sim 50$~G), i.e.\ $f_{\rm cy} \sim 150$~MHz, are already
within the frequency regime of current facilities.

Here we present a search near 150~MHz for radio emission from two
exoplanets with the Giant Metrewave Radio Telescope (GMRT). Previous
radio searches for exoplanets at similar frequencies are discussed
briefly in Section~\ref{s:previous}, and the selection of our
targets is presented in Section~\ref{s:selection}. The GMRT observations
and the data reduction are described in Section~\ref{s:results} along with the
results and conclusions.

\begin{figure}
\centerline{\includegraphics[width=8.5cm]{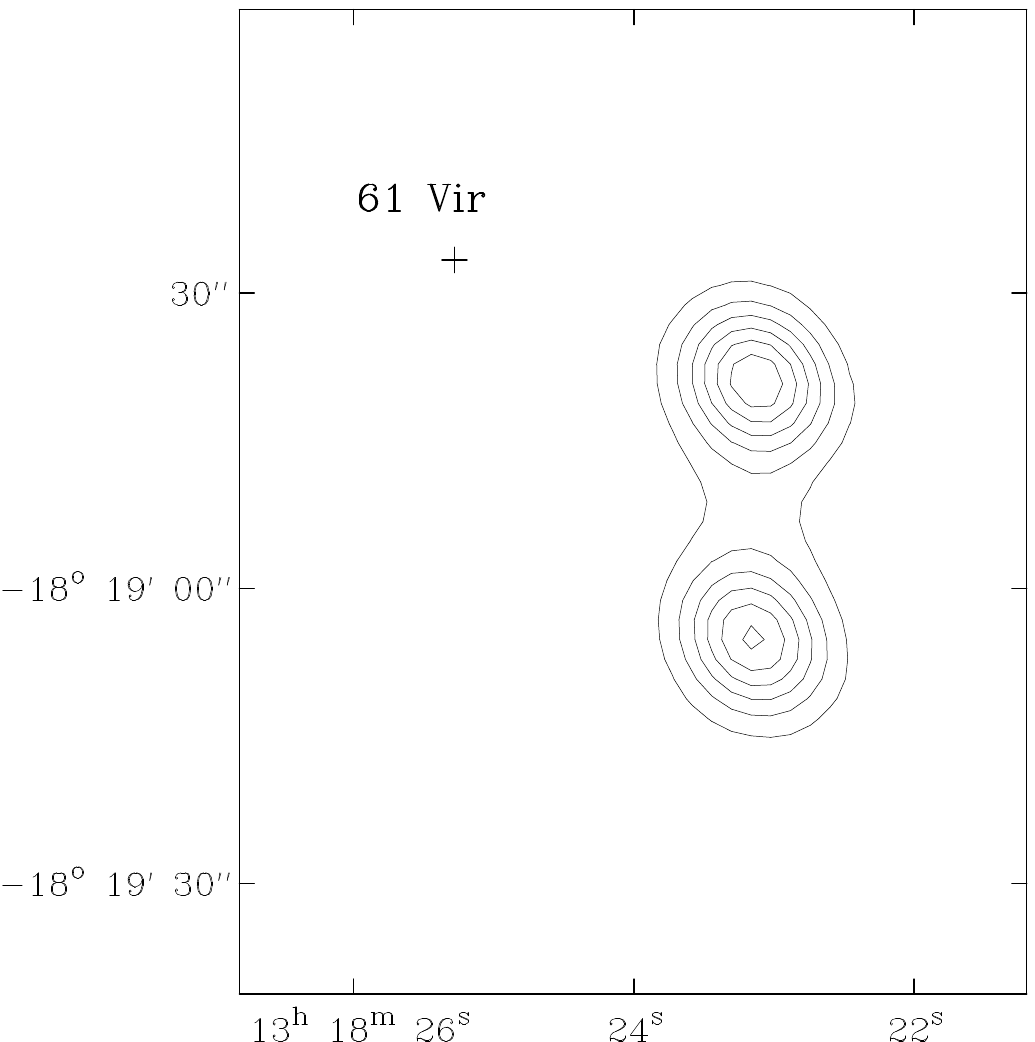}}
\caption{Very Large Array (VLA) image of 61 Vir, at 4.85~GHz, with a
resolution of $12.3 \times 10.0$~arcsec$^2$ at a position angle of
$46^\circ$. The contours are every 1~mJy~beam$^{-1}$. The equatorial
coordinates are J2000.\label{f:VLA}}
\end{figure}

\section{Previous Searches}\label{s:previous}

\citet{2014A&A...562A.108S} searched for radio emission from 175
exoplanets at 150~MHz from the TIFR GMRT Sky Survey (TGSS), which has a
resolution of about 20~arcsec. They did not find evidence of radio
emission associated with their targets, and obtained $3\sigma$ upper
limits of between 8.7 and 136~mJy. \citeauthor{2014A&A...562A.108S} did
note that there was an elongated (about 1~arcmin) radio source at
150~MHz close to 61 Vir, which has three identified exoplanets
\citep{2010ApJ...708.1366V}. This radio source is also seen Northern VLA
Sky Survey (NVSS, \citealt{1998AJ....115.1693C}) at 1.4~GHz, with a
lower resolution of 45~arcsec. \citeauthor{2014A&A...562A.108S} comment
that a higher resolution radio image is needed to resolve whether this
is a chance association. Higher resolution radio observations of this
source are available, as it was observed with the Very Large Array (VLA)
on 1987 February 24, in CD-array at 4.85~GHz (project AC185). We downloaded
these observations from the National Radio Astronomy Observatory
(NRAO) Science Data Archive, and then calibrated and imaged them using
standard procedures in the \textsc{Astronomical Image Processing
System (AIPS)} (e.g.\ \citealt{2003ASSL..285..109G}). The observations
had a 50~MHz bandwidth centred at 4.85~GHz, with about 1 hour spend
observing the source. Figure~\ref{f:VLA} shows an image of these
observations, after primary beam correction. The position of 61 Vir
shown in Fig.~\ref{f:VLA} is for epoch 1987.15, i.e.\ that of the VLA
observations, using the Hipparcos proper motion
\citep{2007A&A...474..653V}. These observations show a compact
(separation about 30~arcsec) radio double source, close but clearly
offset from the position of 61 Vir. This double radio source is
presumably extragalactic, and therefore a chance association with 61
Vir. The flux density of this source at 4.85~GHz is $\approx 16.5$~mJy,
which combined with the TGSS `alternative data release'
\citep{2017A&A...598A..78I} flux density of 229.2 mJy at 150~MHz imply a
radio spectral index of $\alpha$ (here defined in the sense that flux
density $S$ scales with frequency $\nu$ as $S \propto \nu^{-\alpha}$),
of $\approx 0.8$ which is typical of extragalactic radio sources (e.g.\
\citealt{2003A&A...404...57Z, 2010A&A...511A..53V,
2018MNRAS.474.5008D}).

\citet{2015MNRAS.446.2560M} searched for radio emission from 17
exoplanets, using the Murchison Widefield Array (MWA) at 154~MHz. No
targets were detected, with $3\sigma$ upper limits of between 15.2 and
112.5~mJy. Deeper observations at 150~MHz have been made with the GMRT
by \citet{2011A&A...533A..50L, 2013A&A...552A..65L} of HD 189733b, HD
209458b and HAT-P-11b. A possible detections at $3.9 \pm 1.3$~mJy for
was reported for HAT-P-11b, as was a source close to HD 189733b with
$1.9 \pm 0.7$~mJy. (At higher frequencies, \citealt{2009A&A...500L..51L}
report upper limits for HD 189733b from GMRT observations at 240 and 614
MHz.) \cite{2018MNRAS.478.1763L} discuss limits from the MWA circular
polarisation survey at 200~MHz for declinations below $+30^\circ$
\citep{2018MNRAS.478.2835L}, and give $3\sigma$ upper limits from 4.0 to
45.0~mJy for 18 exoplanets. In addition,
\citeauthor{2018MNRAS.478.1763L} present targeted GMRT observation at
150~MHz towards one exoplanet, V830 Tau b, which gives a $3\sigma$ upper
limit of 4.5~mJy. Deeper limits are provided by Low-Frequency Array
(LOFAR) observations at 150~MHz, by \citet{2018A&A...612A..52O}, with
$3\sigma$ upper limits between 0.57 and 0.98 mJy for three exoplanets.

\section{Source Selection}\label{s:selection}

The detectability of cyclotron radio emission from exoplanets is
governed by: (a) the magnitude of the emission, which is a function of
the system parameters (the stellar properties, orbital separation, and
distance to the system), and (b) the gyrofrequency of the emission,
which is solely a function of the planetary magnetic field ($B_{\rm
p}$). Current radio facilities, such as the GMRT, are able to achieve
sensitivities of $\sim 1$~mJy and frequencies as low as 150~MHz. While
$\sim 1$~mJy sensitivities are adequate to detect radio emission from
the nearest close-in exoplanets \citep{2008A&A...490..843J}, the
observable minimum cyclotron frequency of 150~MHz restricts the
observable planetary magnetic field to $B_{\rm p} \gtrsim 50$~G. Radio
searches in the past have typically optimised the first of the two
factors above, i.e.\ targeted nearby short-period exoplanets with the
highest expected emission, while hoping that $B_{\rm p} \gtrsim 50$~G so
that the emission is in the observable frequency range to begin with.
The non-detections from numerous previous searches, despite the high
precisions achieved, raise the question of whether the latter assumption
of $B_{\rm p} \gtrsim 50$~G is actually applicable for the observed
targets.

We aimed to conduct a focused search for potentially radio bright
exoplanets  selected based on factors that could contribute to
their magnetic field strengths. Planetary magnetic fields are
thought to be caused predominantly from `dynamos' in their convective
fluidic interiors. The nature of dynamos in planetary  interiors is
still faced with several open questions \citep{2003E&PSL.208....1S}.
However, some macroscopic dependencies of a planetary magnetic field on
the bulk properties can be understood from first order theory and
empirical trends observed for solar system planets. The key factors
governing the strength of a dynamo in a convective planetary interior
are the extent of its electrically conducting region (the volume and
conductivity of the convective region) and the planetary rotation. The
magnetic moment ($\mu_{\rm p}$) varies as $\mu_{\rm p} \propto \sigma
R_{\rm p}^3 \Omega_{\rm rot}$, where $\sigma$ is the electrical
conductivity of the interior, $R_{\rm p}$ is the planetary radius, and
$\Omega_{\rm rot}$ is the rotation frequency
\citep{2009P&SS...57.1405D}. Magnetic moments of planets in the solar
system follow this behaviour, though the conducting material is
different between giant planets and terrestrial planets. While
metallic hydrogen forms the conducting layer in giant planet interiors,
the conducting layer in the Earth is caused by the liquid Fe core.

For a given planet type (rocky or giant) assuming similar rotation
periods, larger radii and, hence, more extended conductive interiors may
be expected to cause larger magnetic moments as noted above. Thus, the
magnetic field strength at the surface of the planet also increases with
radius \citep{2018haex.bookE...9L}. This is evident from the larger
magnetic moment and magnetic field in Jupiter compared to Saturn. We,
therefore, focused on targets that were giant planets  orbiting nearby
stars with planetary radii similar to or larger
than that of Jupiter ($R_{\rm J}$). With this condition on radius, we also
considered planets with very different masses and, hence, densities, to
span a range in possible metallicities in their interiors. The metallicity
in giant exoplanets is known to increase with decreasing mass
\citep[e.g.,][]{2016ApJ...831...64T, Atreya2018, 2019ARA&A..57..617M, 2019ApJ...887L..20W}.
The metallicities, in turn, could imply different conductivities
in the planetary interior.

We observed two targets motivated by the above criteria. The first was
the hot Saturn WASP-80b \citep{2013A&A...551A..80T} with a mass of 0.54 $M_{\rm J}$ and
radius of 1.00 $R_{\rm J}$ \citep{2015MNRAS.450.2279T}, i.e.\ a density of 0.72 g~cm$^{-3}$
compared to 1.33 g~cm$^{-3}$ for Jupiter. The second target was the hot Jupiter
Qatar-1b \citep{2011MNRAS.417..709A} with a mass of 1.294 $M_J$ and radius of
1.143 $R_{\rm J}$ \citep{2017AJ....153...78C}, i.e.\ a density of 1.15 g~cm$^{-3}$.
Both targets have radii comparable to or larger than Jupiter but densities lower
than Jupiter, and with very different masses and potential metallicities. WASP-80b
is the primary target, as it is at a closer distance than Qatar-1b, $\approx 60$~pc
compared with $\approx 190$~pc.

\section{Results and Conclusions}\label{s:results}

Observations towards Qatar-1b and WASP80b were made with the GMRT, which
consists of 30 antennas each 45~m in diameter (see \citealt{2002IAUS..199..439R}),
in 2015 September. These were scheduled to cover a secondary eclipse of the
exoplanet\footnote{Based on timing from the NASA Exoplanet archive at:
\url{http://exoplanetarchive.ipac.caltech.edu/}.}. The aim was to look
for emission associated with the exoplanet, and if detected, then to look
for decrease in emission during the eclipse to confirm the exoplanet
emission. The observations were made with a bandwidth of 16.7~MHz,
centred at 147.7~MHz, using 256 channels. Both left and right circular
polarisations were observed. However, strong interference away from
the centre of band -- particularly for the Qatar-1b observations --
meant than in practice smaller bandwidths of 5.2 and 14.0~MHz, centred
at 145.6 and 147.8~MHz were used for Qatar-1b and WASP80b
respectively. For each observing run scans on 3C286 and 3C48 were included
at the beginning and end the run, for flux density scale calibration, and
nearby compact sources were observed every 25 min or so, to monitor the
amplitude of phase calibration of the antennas through the run.

The observed $u,v$-data were processed using standard procedures in
AIPS. Interference was flagged by eye, and then several channels near
the centre of the band were collapsed together, and these data were tied
to the flux scale of \citet{2012MNRAS.423L..30S} from the observations
of 3C48 and 3C286. The data were then corrected for antenna-based
amplitude and phase variations through the run. The calibration from
these collapsed data was then applied to the wider bandwidth using a
antenna based bandpass calibration, from the observations of 3C286 and
3C48.

\begin{figure}
\centerline{\includegraphics[width=8.5cm]{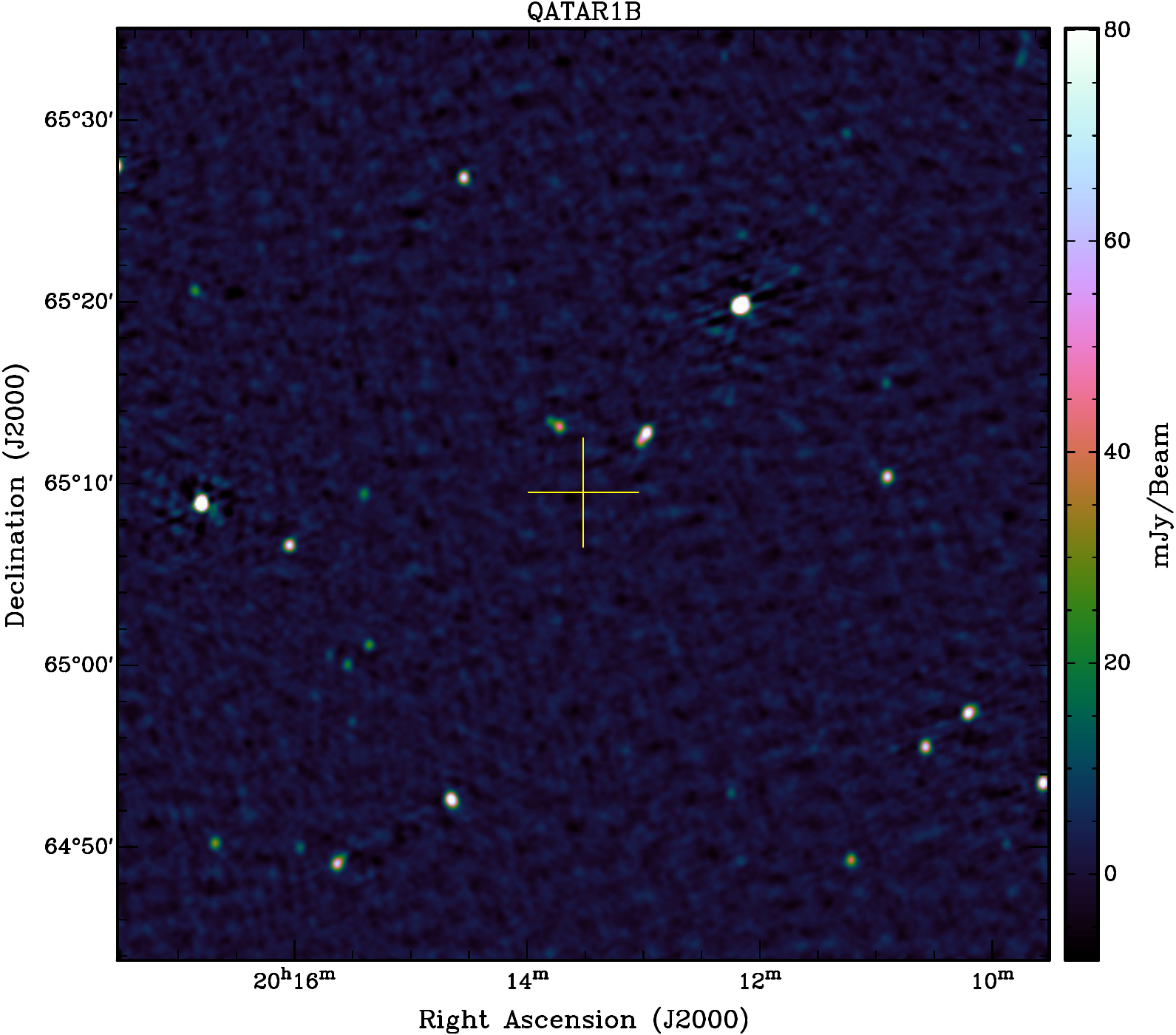}}
\caption{Image of Qatar-1b made with the GMRT at 145.6~MHz. The position
of Qatar-1b is indicated by the cross.\label{f:qatar}}
\end{figure}

\begin{figure}
\centerline{\includegraphics[width=8.5cm]{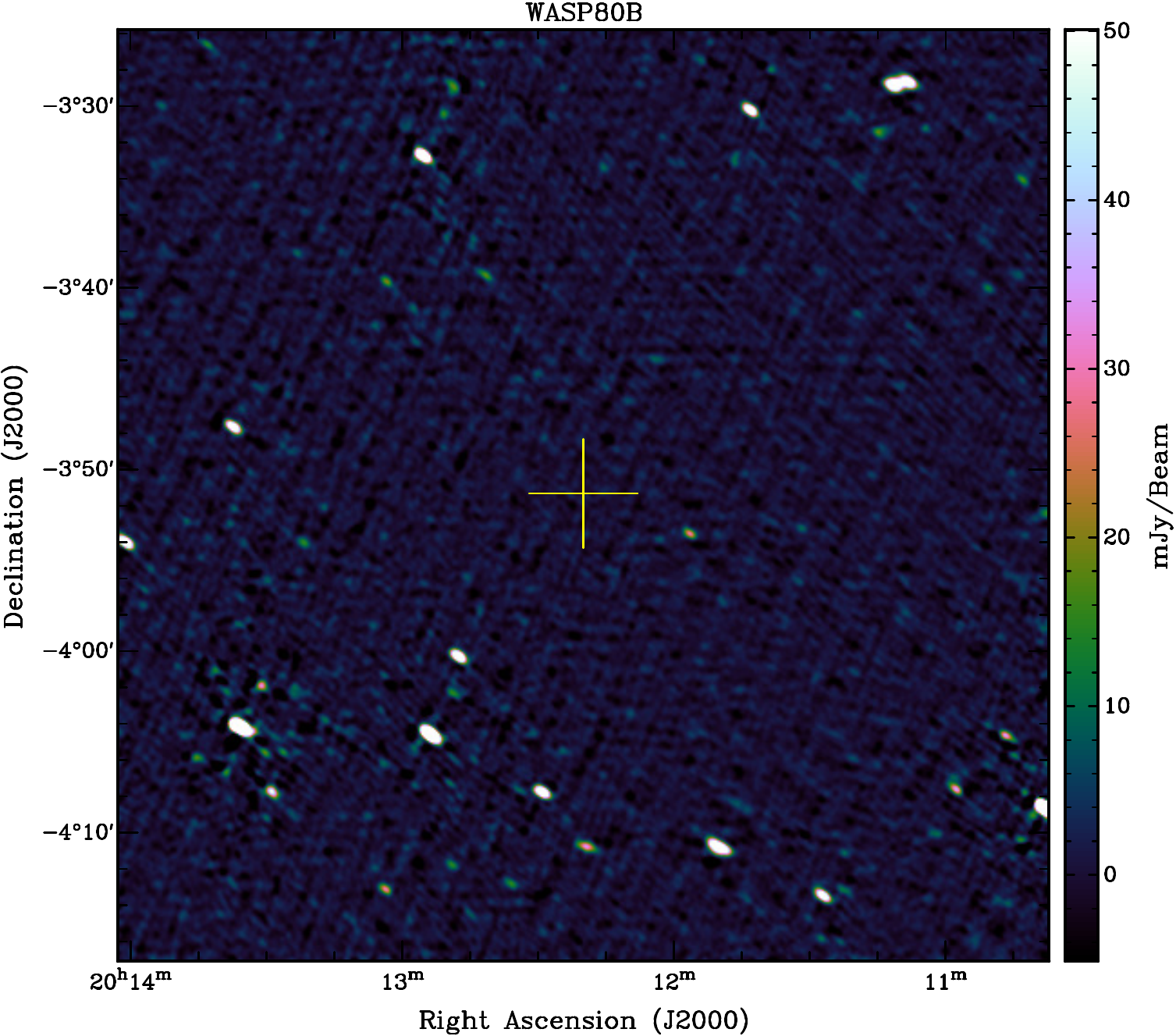}}
\caption{Image of WASP-80b made with the GMRT at 147.8~MHz. The position
of WASP-80b is indicated by the crosses.\label{f:wasp}}
\end{figure}

The calibrated $u,v$-data were imaged using multiple `facets', which is
needed due to the relatively large field-of-view of the observations
(the HPBW of the primary beam is about $2$~deg at 150~MHz). Both
circular polarisations were combined to make Stokes I images.
Subsequently several iterations of self-calibration were applied, to
correct for residual calibration errors. Phase only self-calibration was
applied on decreasing time intervals down to 2~min, with a final
self-calibration applied for amplitude and phase on a timescale of
10~min, ensuring that the overall amplitude scaling was preserved.

We do not detect any radio emission from either of our targets The central
portions of the final images of Qatar-1b and WASP-80b are
shown in Figs~\ref{f:qatar} and \ref{f:wasp}. These were made using
`natural' weighting, for maximum sensitivity. The r.m.s.\ sensitivities
are 1.8 and 1.5 mJy~beam$^{-1}$ for Qatar-1b and WASP80b respectively.
No emission is detected from the positions of the exoplanets. Given the
observations span secondary eclipses, any radio emission from the
exoplanets would be excluded for some of the observations. The total time
on source was 6.3 and 7.7~hr for Qatar-1b and WASP80b respectively, of
which 2.0 and 1.1~hr corresponded to the eclipses.
Taking these timings into account, the $3\sigma$ limits on any radio
emission from  Qatar-1b and WASP80b are 5.9 and 5.2~mJy, at 145.6 and
147.8~MHz respectively.

As discussed in Section~\ref{s:previous}, \cite{2014A&A...562A.108S}
noted a radio source in the TGSS close to 61 Vir, which has three
planets. However, this radio source is a double source, typical of
extragalactic sources, and is significantly offset from the stellar
position. We have cross-matched a recent version\footnote{From
\url{http://exoplanet.eu/catalog/}, with 3221 planetary systems (as of
2020 October 8).} of an exoplanet catalogue (see
\citealt{2011A&A...532A..79S}) with the TGSS ADR1 source catalogue from
\cite{2017A&A...598A..78I}. There are four TGSS sources that closely
($<10$~arcsec) match exoplanets: three are with radio pulsars which have
exoplanets (namely PSR B0329$+$54, PSR B1957$+$20 and PSR B0943$+$10),
with the other match being with Kepler-652b. The positional offset
between the TGSS source and Kepler-652b is $\sim 2$~arcsec, which is
comparable to the error in the TGSS source position. Although a chance
association this close is not very probable, it does not seem likely
this is a radio exoplanet detection. The TGSS source is quite bright,
with a flux density of $69$~mJy at 150~MHz, which is much brighter than
what might be expected from an exoplanet, especially given Kepler-652b
is a Neptune like planet, and is relatively distant ($\sim 0.42$~kpc).
The TGSS source is also detected in the Westerbork Northern Sky Survey
(WENSS; \citealt{1997A&AS..124..259R}) at 325 MHz with $~50$~mJy, and at
1.4 GHz in the NVSS survey (\citeauthor{1998AJ....115.1693C}) with
$~27$~mJy. These show it has a radio spectral index ($\alpha \approx
0.4$), consistent with an extragalactic radio source, and is considered
a chance association with Kepler-652b.

The limits provided here for Qatar-1b and WASP-80b are more sensitive than
those available from wide-field radio surveys at similar frequencies
(e.g.\ a median $3\sigma$ upper limit of about 25~mJy at 151~MHz from
\citealt{2014A&A...562A.108S}). Considering that our chosen targets
are giant exoplanets, with potentially higher detectability than smaller
planets, our non-detections call for future observations with even higher
sensitivities and, where possible, lower frequencies. It is evident from
our present, and numerous previous searches, that single eclipse observations
may not be adequate for such detections. Future observations may consider
co-adding eclipse observations over multiple epochs to improve the
sensitivities, as is often pursued to detect low-amplitude spectral
features of transiting planets in the near-infrared \citep[e.g.,][]{2014Natur.505...69K,
2014Sci...346..838S}. The increasing number of exoplanets detected
around nearby stars may also make future observations of their radio
emission more feasible.

\section*{Acknowledgements}

We thank the staff of the GMRT that made these observations possible.
The GMRT is run by the National Centre for Radio Astrophysics of the
Tata Institute of Fundamental Research. DAG thanks the Science and
Technology Facilities Council for financial support for these
observations. The National Radio Astronomy Observatory is a facility of
the National Science Foundation operated under cooperative agreement by
Associated Universities.

\section*{Data Availability}

The VLA observations used for Fig.~\ref{f:VLA} are available in the NRAO
Science Data Archive at \url{https://archive.nrao.edu/}, for Project
Code AC185, and GMRT observations used for Figs \ref{f:qatar} and
\ref{f:wasp} are available in the GMRT Online Archive at
\url{https://naps.ncra.tifr.res.in/goa/}, for Proposal Code 28\_065.

\section*{Note Added In Proof}

The VLA observations of Vir 61 from 1987 have previously been
published in \citet{2012MNRAS.424.1206W}.

%

\label{lastpage}


\begin{thebibliography}{}

\bibitem[Alsubai et al.(2011)Alsubai et al.]{2011MNRAS.417..709A}
  Alsubai K.~A., et al.,
    2011, MNRAS, 417, 709

\bibitem[Atreya et al.(2018)Atreya et al.]{Atreya2018}
  Atreya, S., Crida, A., Guillot, T., Lunine, J., Madhusudhan, N., Mousis, O.,
    2018, in Baines K., Flasar F., Krupp N., Stallard T., eds,
    Saturn in the 21st Century. Cambridge University Press, p.~5.

\bibitem[Collins et al.(2017)Collins, Kielkopf \& Stassun]{2017AJ....153...78C}
  Collins K.~A., Kielkopf J.~F., Stassun K.~G.,
    2017, AJ, 153, 78

\bibitem[Condon et al.(1998)Condon et al.]{1998AJ....115.1693C}
  Condon J.~J., Cotton W.~D., Greisen E.~W., Yin Q.~F., Perley R.~A.,
  Taylor G.~B., Broderick J.~J., 1998, AJ, 115, 1693

\bibitem[de Gasperin et al.(2018)de Gasperin, Intema \& Frail]{2018MNRAS.474.5008D}
  de Gasperin F., Intema H.~T., Frail D.~A.,
    2018, MNRAS, 474, 5008

\bibitem[Deming \& Seager(2017)Deming \& Seager]{2017JGRE..122...53D}
  Deming L.~D., Seager S.,
    2017, J.\ Geophys.\ Res.: Planets, 122, 53

\bibitem[Durand-Manterola(2009)Durand-Manterola]{2009P&SS...57.1405D}
  Durand-Manterola H.~J.,
    2009, Planet. \& Space Sci., 57, 1405

\bibitem[Greisen(2003)Greisen]{2003ASSL..285..109G}
  Greisen E.~W.,
    2003, in Heck A., ed., Information Handling in Astronomy -- Historical
    Vistas, Astrophysics and Space Science Library, Vol.\ 285, Kluwer
    Academic Publishers, p.~109

\bibitem[Grie{\ss}meier et al.(2007)Grie{\ss}meier, Zarka \& Spreeuw]{2007A&A...475..359G}
  Grie{\ss}meier J.-M., Zarka P., Spreeuw H.,
    2007, A\&A, 475, 359

\bibitem[Intema et al.(2017)Intema et al.]{2017A&A...598A..78I}
  Intema H.~T., Jagannathan P., Mooley K.~P., Frail D.~A.,
    2017, A\&A, 598, A78

\bibitem[Jardine \& Collier Cameron(2008)Jardine \& Collier Cameron]{2008A&A...490..843J}
  Jardine M., Collier Cameron A.,
    2008, A\&A, 490, 843

\bibitem[Kreidberg et al.(2014)Kreidberg et al.]{2014Natur.505...69K}
 Kreidberg L., et al.,
 2014, Nature, 505, 69

\bibitem[Kreidberg(2018)Kreidberg]{2018haex.bookE.100K}
  Kreidberg L.,
    2018, in Deeg H.~J., Belmonte J.~A., eds., Handbook of Exoplanets.
    Springer Nature, p.~2083

\bibitem[Lazio(2018)Lazio]{2018haex.bookE...9L}
  Lazio T.~J.~W.,
    2018, in Deeg H.~J., Belmonte J.~A., eds., Handbook of Exoplanets.
    Springer Nature, p.~817

\bibitem[Lazio et al.(2004)Lazio et al.]{2004ApJ...612..511L}
  Lazio T.~J., W., Farrell W.~M., Dietrick J., Greenlees E., Hogan E., Jones C., Hennig L.~A.,
    2004, ApJ, 612, 511

\bibitem[Lecavelier Des Etangs et al.(2009)Lecavelier Des Etangs et al.]{2009A&A...500L..51L}
  Lecavelier Des Etangs A., Sirothia S.~K., Gopal-Krishna, Zarka P.,
    2009, A\&A, 500, L51

\bibitem[Lecavelier Des Etangs et al.(2011)Lecavelier Des Etangs et al.]{2011A&A...533A..50L}
  Lecavelier Des Etangs A., Sirothia S.~K., Gopal-Krishna, Zarka P.,
    2011, A\&A, 533, A50

\bibitem[Lecavelier des Etangs et al.(2013)Lecavelier des Etangs et al.]{2013A&A...552A..65L}
  Lecavelier des Etangs A., Sirothia S.~K., Gopal-Krishna, Zarka P.,
    2013, A\&A, 552, A65

\bibitem[Lenc et al.(2018)Lenc et al.]{2018MNRAS.478.2835L}
  Lenc, E., Murphy, T., Lynch, C. R., Kaplan, D. L., Zhang, S. N.,
    2018, MNRAS, 478, 2835

\bibitem[Lynch et al.(2018)Lynch et al.]{2018MNRAS.478.1763L}
  Lynch, C. R., Murphy, T., Lenc, E., Kaplan, D. L.,
    2018, MNRAS, 478, 1763

\bibitem[Madhusudhan(2019)Madhusudhan]{2019ARA&A..57..617M}
  Madhusudhan N.,
    2019, ARA\&A, 57, 617

\bibitem[Murphy et al.(2015)Murphy et al.]{2015MNRAS.446.2560M}
  Murphy T., et al.,
    2015, MNRAS, 446, 2560

\bibitem[Nichols(2011)Nichols]{2011MNRAS.414.2125N}
  Nichols J.~D.,
    2011, MNRAS, 414, 2125

\bibitem[Nichols(2012)Nichols]{2012MNRAS.427L..75N}
  Nichols J.~D.,
    2012, MNRAS, 427, L75

\bibitem[Noyola et al.(2014)Noyola, Satyal \& Musielak]{2014ApJ...791...25N}
  Noyola J.~P., Satyal S., Musielak Z.~E.,
    2014, ApJ, 791, 25

\bibitem[O'Gorman et al.(2018)O'Gorman et al.]{2018A&A...612A..52O}
  O'Gorman, E., Coughlan, C. P., Vlemmings, W., Varenius, E., Sirothia, S., Ray, T. P., Olofsson, H.,
    2018, A\&A, 612, A52

\bibitem[Rao(2002)Rao]{2002IAUS..199..439R}
  Rao A.~P.,
    2002, in Rao A.~P., Swarup G., Gopal-Krishna, eds, Proc. IAU Symp.\ 199,
    The Universe at Low Radio Frequencies. Astron.\ Soc.\ Pac., San Francisco, p.~439

\bibitem[Rengelink et al.(1997)Rengelink et al.]{1997A&AS..124..259R}
  Rengelink R.~B., Tang Y., de Bruyn A.~G., Miley G.~K., Bremer M.~N., R\"ottgering H.~J.~A., Bremer M.~A.~R.,
    1997, A\&AS, 124, 259

\bibitem[Scaife \& Heald(2012)Scaife \& Heald]{2012MNRAS.423L..30S}
  Scaife A.~M.~M., Heald G.~H.,
    2012, MNRAS, 423, L30

\bibitem[Schneider et al.(2011)Schneider et al.]{2011A&A...532A..79S}
  Schneider J., Dedieu C., Le Sidaner P., Savalle R., Zolotukhin I.,
    2011, A\&A, 532, A79

\bibitem[Sirothia et al.(2014)Sirothia et al.]{2014A&A...562A.108S}
  Sirothia S.~K., Lecavelier des Etangs A., Gopal-Krishna, Kantharia
  N.~G., Ishwar-Chandra C.~H.,
    2014, A\&A, 562, A108

\bibitem[Stevenson(2003)Stevenson]{2003E&PSL.208....1S}
  Stevenson D.~J.,
    2003, Earth \& Planet. Sci. Lett., 208, 1

\bibitem[Stevenson et al.(2014)Stevenson et al.]{2014Sci...346..838S}
 Stevenson K.~B., et al.,
 2014, Science, 346, 838

\bibitem[Thorngren et al.(2016)Thorngren et al.]{2016ApJ...831...64T}
  Thorngren D.~P., Fortney J.~J., Murray-Clay R.~A., Lopez E.~D.,
    2016, ApJ, 831, 64

\bibitem[Triaud et al.(2013)Triaud et al.]{2013A&A...551A..80T}
  Triaud A.~H.~M.~J., et al.,
    2013, A\&A, 551, A80

\bibitem[Triaud et al.(2015)Triaud et al.]{2015MNRAS.450.2279T}
  Triaud A.~H.~M.~J., et al.,
    2015, MNRAS, 450, 2279

\bibitem[van Leeuwen(2007)van Leeuwen]{2007A&A...474..653V}
  van Leeuwen F.,
    2007, A\&A, 474, 653

\bibitem[Vogt et al.(2010)Vogt et al.]{2010ApJ...708.1366V}
  Vogt S.~S., et al.,
    2010, ApJ, 708, 1366

\bibitem[Vollmer et al.(2010)Vollmer et al.]{2010A&A...511A..53V}
  Vollmer B., et al.,
    2010, A\&A, 511, A53

\bibitem[Welbanks et al.(2019)Welbanks et al.]{2019ApJ...887L..20W}
  Welbanks L., Madhusudhan N., Allard N.~F., Hubeny I., Spiegelman F.,
  Leininger T.,
    2019, ApJ, 887, L20

\bibitem[Wyatt et al.(2012)Wyatt et al.]{2012MNRAS.424.1206W}
  Wyatt M.~C., et al.,
    2012, MNRAS, 424, 1206

\bibitem[Zarka(2007)Zarka]{2007P&SS...55..598Z}
  Zarka P.,
    2007, Planet. \& Space Sci., 55, 598

\bibitem[Zarka et al.(2001)Zarka et al.]{2001Ap&SS.277..293Z}
  Zarka P., Treumann R.~A., Ryabov B.~P., Ryabov V.~B.,
    2001, Ap\&SS, 277, 293

\bibitem[Zhang et al.(2003)Zhang et al.]{2003A&A...404...57Z}
  Zhang X., Reich W., Reich P., Wielebinski R.,
    2003, A\&A, 404, 57

\end{thebibliography}
\end{document}